\documentclass[a4paper,11pt]{article}
\usepackage[utf8]{inputenc}
\usepackage{jcappub} 
\usepackage{graphicx,rotating,wrapfig,float}
\usepackage{tabularx}
\usepackage{multirow}

\newcommand{\alp}{\ensuremath{\alpha}}
\newcommand{\bet}{\ensuremath{\beta}}
\newcommand{\gam}{\ensuremath{\gamma}}
\newcommand{\ardm}{ArDM}

\newcommand{\ardmrI}{ArDM\,Run\,I}

\newcommand{\kr}{\ensuremath{\,\mathrm{^{83m}Kr}}}
\newcommand{\ar}{\ensuremath{\,\mathrm{^{39}Ar}}}
\newcommand{\co}{\ensuremath{\,\mathrm{^{57}Co}}}
\newcommand{\rn}{\ensuremath{\,\mathrm{^{222}Rn}}}
\newcommand{\cf}{\ensuremath{\,\mathrm{^{252}Cf}}}
\newcommand{\ur}{\ensuremath{\,\mathrm{^{238}U}}}
\newcommand{\tho}{\ensuremath{\,\mathrm{^{232}Th}}}
\newcommand{\pot}{\ensuremath{\,\mathrm{^{40}K}}}
\newcommand{\cob}{\ensuremath{\,\mathrm{^{60}Co}}}

\newcommand{\kev}{\ensuremath{\,\mathrm{keV}}}
\newcommand{\mev}{\ensuremath{\,\mathrm{MeV}}}

\newcommand{\msq}{\ensuremath{\,\mathrm{m^{2}}}}
\newcommand{\msqps}{\ensuremath{\,\mathrm{m^{2}s^{-1}}}}

\newcommand{\mus}{\ensuremath{\,\mu\mathrm{s}}}
\newcommand{\ns}{\ensuremath{\,\mathrm{ns}}}

\newcommand{\hz}{\ensuremath{\,\mathrm{Hz}}}

\title{Backgrounds and pulse shape \\ 
discrimination in the ArDM \\
liquid argon TPC}

\collaboration{ArDM Collaboration}

\author[a]{J.~Calvo}
\author[a]{C.~Cantini}
\author[a]{P.~Crivelli}
\author[b]{M.~Daniel}
\author[a]{S.~Di~Luise}
\author[a]{A.~Gendotti}
\author[a]{S.~Horikawa}
\author[a]{L.~Molina-Bueno}
\author[b]{B.~Montes}
\author[a]{W.~Mu}
\author[a]{S.~Murphy}
\author[a]{G.~Natterer}
\author[a]{K.~Nguyen}
\author[a]{L.~Periale}
\author[a]{Y.~Quan}
\author[a]{B.~Radics}
\author[a]{C.~Regenfus}
\author[b]{L.~Romero}
\author[a]{A.~Rubbia}
\author[b]{R.~Santorelli}
\author[a]{F.~Sergiampietri}
\author[a]{T.~Viant}
\author[a]{S.~Wu}

\affiliation[a]{ETH Zurich, Institute for Particle Physics, Zurich, Switzerland}
\affiliation[b]{CIEMAT, Div. de F{\'\i}sica de Particulas, Avda. Complutense, 22, E-28040, Madrid, Spain}

\emailAdd{andre.rubbia@cern.ch}

\abstract{The \ardm\ experiment completed a single-phase commissioning run (\ardmrI) in 2015 with an active liquid argon target of nearly one tonne in mass. The analysis of the data and comparison to simulations allowed for a test of the crucial detector properties and confirmed the low background performance of the setup. The statistical rejection power for electron recoil events using the pulse shape discrimination method was estimated using data from a \cf\ neutron calibration source. Electron and nuclear recoil band profiles were found to be well described by Gaussian distributions. Employing such a model we derive values for the electron recoil statistical rejection power of more than 10$^8$ in the tonne-scale liquid argon target for events with more than 50 detected photons at a $50$\% acceptance for nuclear recoils. The \rn\ emanation rate of the ArDM cryostat at room temperature was found to be 65.6$\pm$0.4\,$\mu$\hz/l, and the \ar\ specific activity from the employed atmospheric argon to be 0.95$\pm$0.05\,Bq/kg. The cosmic muon flux at the Canfranc underground site was determined to be between 2 and 3.5$\times 10^{-3}$\msqps . These results pave the way for the next physics run of \ardm\ in the double-phase operational mode.}


\begin{document}

\maketitle

\section{Introduction}
\ardm\ is a direct dark matter detection experiment for Weakly Interacting Massive Particles (WIMPs) searches~\cite{Rubbia:2005ge}. The detector system consists of a cylindrical Time Projection Chamber (TPC) containing a liquid argon (LAr) target mass of around 850\,kg. The \ardm\ experiment is able to detect signals produced by elastic scattering, from WIMPs or neutrons producing nuclear recoils (NR) and from background particles like \bet\ or \gam\ producing electron recoils (ER), on argon atoms in the active volume.

In 2015, a series of commissioning runs with gaseous or liquid argon targets have been performed in single-phase mode to explore the functionality and performance of the detector. The detector was first commissioned at room temperature with a gaseous argon (GAr) target to verify the basic functionality of the detector components. Then the detector was cooled down and operated with a cold GAr target to test the cryogenic system and the light detection system at low temperature. Afterwards, data taking with a full LAr target was performed. At last, the detector was pumped and refilled with GAr to investigate the detector stability. A detailed description of the detector and the commissioning runs can be found in Ref.\,\cite{1475-7516-2017-03-003}.

The various background sources of the detector have been studied using data collected during \ardmrI\ and compared with Monte-Carlo (MC) simulations. At low energy (below 1 MeV), the background originates from \ar-\bet\ decays, radioactive contaminants in the materials of the detector components, and from external \gam\ sources. At medium energy (1-5 MeV), the background originates from Rn and its progeny, emanated from the small contamination of \ur\ and \tho\ in the detector materials. In addition, at high energy ($>$ 5 MeV) the ArDM detector also recorded cosmic muon events.

The time structure of LAr scintillation is strongly correlated with the nature of the interaction, which provides a way, the pulse shape discrimination (PSD) method, to reject the ER, or so-called e-like, background events~\cite{Boulay:2006mb, Benetti:2007cd, Agnes:2014bvk, Amaudruz:2016qqa}. Analyzing data taken with a \cf\ calibration source, the band profiles of ER or NR events were found to be well described by Gaussian distributions. Using the separation between the ER and NR bands, the capability of the PSD method to reject ER events was demonstrated.

This paper is organized as follows. In Section\,\ref{sec:lebkg}, the low energy ER background is discussed using the comparison between the data taken with LAr target and the MC simulations of different radioactive isotopes, including results on the measurement of the \ar\ activity. In Section\,\ref{sec:rn}, results are presented on the measurement of the \rn\ emanation using data taken with GAr target at room temperature. Section\,\ref{sec:Muon} discusses cosmic muon rates observed in \ardmrI\; LAr data.  The PSD method and the statistical rejection power results are presented in Section\,\ref{sec:elike}. Finally, the work is concluded in Section\,\ref{sec:conc}.

\section{The ArDM experiment}
\label{experiment}

The Argon Dark Matter (ArDM) experiment (LSC-EXP-08, CERN-RE18) is a particle physics experiment at the Spanish underground site LSC (Laboratorio Subterr\'{a}neo de Canfranc) to perform direct searches for WIMP-induced nuclear recoils. The ArDM detector is designed as a tonne-scale dual-phase liquid-argon time projection chamber (LAr TPC) to detect elastic scattering of WIMPs on argon nuclei, by observing ionisation and scintillation events, which are produced by the recoiling nucleus in the argon medium \cite{Rubbia:2005ge}. 
During the first years of the project, the experiment was first developed and constructed at CERN \cite{Amsler:2010yp,Marchionni:2010fi} and then installed at LSC in 2014 \cite{Badertscher:2013ygt}. In 2015 ArDM completed a first data taking period of more than six months in the single-phase operational mode with a full LAr target of 850\,kg (ArDM Run I) \cite{1475-7516-2017-03-003}. In 2016 the experiment was upgraded for the a dual-phase operation. 

The ArDM LAr TPC consists of 
24 low-radioactivity cryogenic 8'' PMTs distributed in two equal arrays for light readout, the top PMT array above the LAr target in the gaseous phase and the bottom array immersed in LAr. 
Nuclear recoils induced by WIMPs or neutrons, or electron recoils by $\gamma$ radiation, as well as charged particle ($\alpha$, $\beta$ or muons) interactions in the argon medium generate scintillation light (S1) and electron-ion pairs. 
When the detector works in the dual-phase (liquid and gaseous) mode, the electrons can be separated from their ions in an electric field and drift upwards to the argon surface. After being extracted from the LAr to the gaseous argon (GAr) on top, these electrons are accelerated and the secondary scintillation light (S2), which is proportional to the amount of electrons extracted, is produced. Both S1 and S2, which are vacuum ultraviolet (VUV) light with a wavelength around 127 nm, can be wavelength shifted to visible range by a layer of tetraphenylbutadiene (TPB) deposited on the inner surfaces and read out by the PMT arrays. 
In the single-phase commissioning Run I, with zero electric field, only S1 signals were recorded. 
A full detail about the experimental setup of ArDM during Run I is reported in Ref. \cite{1475-7516-2017-03-003}.

Table \ref{table:summary_runstage} summarises the Run-I data recorded at different stages of the operation. 
The detector was commissioned first with a gaseous argon (GAr) target at room temperature, hereafter called warm gas. Then it was cooled down and was operated with a GAr target at LAr temperature, 87\,K (cold gas). These were succeeded by the filling of the ArDM cryostat with a total of $\sim$2 tonnes of LAr. The detector was then operated over six months in the full target mode. 
The full-target data consists of two parts. 
Initially the top part of the polyethylene (PE) neutron shield, entirely surrounding the detector vessel, was left open (open-shield configuration) to allow accesses to various service ports located on the top cover of the vessel. When the stable performance of the detector had been verified the shield was closed completely (closed-shield configuration), resulting in a reduction of $\gamma$ backgrounds originating from the lab environment, which represented the designed configuration of the experiment. After 3.3 billion triggers had been recorded with the full target, the liquid was evaporated slowly and the cryostat was warmed up. 
The data taking was continued over the entire period, including transition stages, such as cool-down, filling or warm-up, that allowed to monitor the functionality of the detector, as well as the purity of the LAr target, at different run stages. 
A total of about 4.7 billion triggered events have been recorded over the period of Run I. 
About 10\% of this total data was collected during calibration runs with radioactive test sources. 
The calibration data collected with the full LAr target is summarized in Table \ref{table:summary_source}. 
Metastable $^{83 \rm m}$Kr atoms, which were injected into the gaseous phase and then diluted in the LAr target, served as a main source of the light yield calibration, along with an external $^{57}$Co source. $^{252}$Cf neutron source was deployed to study the PSD capability of the ArDM detector, as reported below in Section \ref{sec:elike}. 

\begin{table}[htb]
\begin{center}
\begin{tabular}{l@{\extracolsep{\fill}}r} 
\hline
\bf Run stage			& \bf Recorded events [$10^6$]	\\
\hline
Warm gas				& 213	\\
Cool-down 			& 20 		\\ 
Cold gas				& 351 	\\
LAr filling				& 495 	\\
\hline
\bf Full LAr target 		& 		\\
Open shield			& 1286 	\\
Closed shield			& 2025  	\\
\bf Subtotal 			& \bf 3310	\\
\hline
Warm-up	\& warm gas 	& 308 	\\
\hline
\bf Total 				& \bf 4697	\\
\hline
\end{tabular}
\end{center}
\caption{Summary of the number of recorded events at different stages of ArDM Run I.}
\label{table:summary_runstage}
\end{table}

\begin{table}[htb]
\begin{center}
\begin{tabular}{lrrrrrr} 
\hline
\small \bf Source					& $^{83{\rm m}}$Kr 	& $^{57}$Co	& $^{252}$Cf	& $^{60}$Co	& $^{22}$Na	& Total 	\\
\hline 
\small \bf Recorded events [$10^6$] 	& 201 			& 89 			& 48 			& 6			& 6			& 350	\\
\hline 
\end{tabular}
\caption{Number of events recorded with calibration sources in the full LAr target during ArDM Run I.}
\label{table:summary_source}
\end{center}
\end{table}

\section{Analysis of the ER background in ArDM}
\label{sec:lebkg}
For direct WIMP dark matter detection the understanding of the \bet\ or \gam\ induced ER background events is crucial. 
We have produced MC simulations of \gam\ background events from both internal and external sources, as well as for \bet\ background events from \ar\ decay, and compared simulation results with data, leading to an agreement at a $\sim 7\%$ level.

\subsection{Data selection}
\label{sec:lar}
The \ardm\; light detection system uses 12-12 photomultiplier tubes (PMTs) arranged at the top and the bottom of a cryogenic vessel, respectively. The energy of the registered events is characterised by the total light in units of photo-electrons ($pe$). The total light of an event is calculated by finding and summing clusters of photon signals over the data acquisition time window of 4\mus. The light signals, $L_{\rm top}$ and $L_{\rm btm}$, are calculated from the sum of all signal clusters found in the top and the bottom PMTs, respectively. The total detected light $L_{\rm tot}$ is calculated from the sum of the signal from both PMT arrays. The light yield per \kev\ electron-equivalent energy deposit (also denoted as keV$_{\rm ee}$) in the \ardm\ detector was calibrated using \kr\ and \co\ sources during \ardmrI. The obtained light yield was 1.1\,$pe$/\kev\ in LAr. 

A pulse shape discriminator value, $f90$, defined as the ratio of the light detected in the first 90\ns\ of the event to the total detected light, is used during the analysis. This value is used to separate between ER and NR events. In addition, events with extreme $f90$ values ($f90 >$ 0.99 or $f90 < 0.01$) are rejected due to their origin being noise. ER events can be separated by selecting events requiring $f90 < 0.6$.

\begin{figure}[hbt]
\centering
\includegraphics[width=0.95\textwidth]{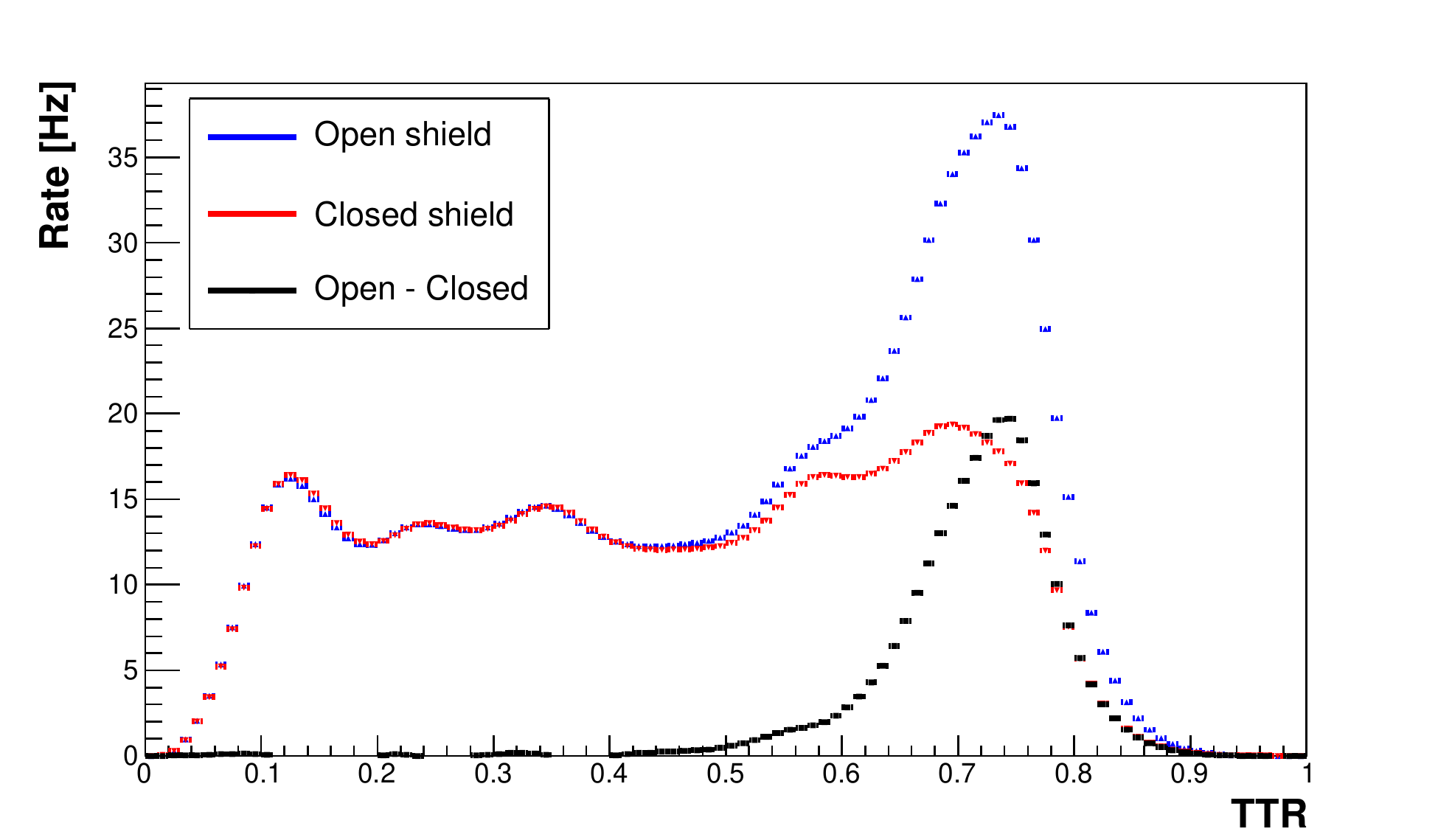}
\caption{$TTR$ distribution during data taken with open (blue), closed (red) top shield configurations. The difference of the two distributions is shown in black. $TTR$ values larger than 0.5 correspond to vertical positions closer to the top of the detector.}
\label{fig:diff1}
\end{figure}

The ratio of the light detected by the top PMT array to the total detected light, called top-to-total ratio, $TTR = L_{\rm top}/L_{\rm tot}$, is used as a discriminator to select events along the vertical position ($Z$-axis in the ArDM coordinate system) in the LAr volume. 
For multiple-scatter events, produced by e.g. a $\gamma$ photon that undergoes several Compton scatterings, this value shows the mean vertical position of the interaction vertices weighted by the energy deposit at each vertex. 
Larger $TTR$ values correspond to vertices closer to the top of the detector. Figure\,\ref{fig:diff1} illustrates the distribution of $TTR$ values in the data.

\subsection{Simulation}
\label{sec:simulation}

A Geant4-based detector simulation software with light ray-tracing has been developed for ArDM. 
The ArDM TPC, the stainless-steel detector vessel and the surrounding PE neutron shield are modeled with approximate geometries and material compositions. 
The response of the ArDM detector results from complex processes involving VUV and visible light. 
The existing Geant4 scintillation and wavelength-shifting processes were tuned to describe different time structures of the VUV light emission for electron and nuclear recoil events in liquid and gaseous argon, and  in the TPB layer properties on the various surfaces, respectively. 
Light propagation (attenuation and Rayleigh scattering) in the LAr medium and boundary processes (reflection and refraction) across different materials are implemented as a function of the wavelength. 
The PMT response to the detected photoelectrons, e.g. pulse shape and amplitude fluctuations, are simulated. 
The simulated events include digitised waveforms of the 24 PMT channels, recorded in the same format as in the ArDM front-end electronics, allowing to reconstruct the events using the same analysis software as used for the experimental data. 
Parameters describing the optical processes were tuned to the data obtained from Run I. 
Measured light spectra for ER events originating from $^{39}$Ar $\beta$-decays and the de-excitation of $^{83{\rm m}}$Kr calibration source, as a function of the vertical position of the interaction vertex reconstructed by the $TTR$ variable, were used for this purpose for their well-understood origins and spatially uniform distributions in the LAr bulk. 
A set of parameters best describing the observed data was evaluated using a Bayesian likelihood technique. 
A detailed description of the ArDM detector simulation software, as well as the tuning procedures, is given in Ref. \cite{Calvo:2016nwp}.

To fully understand the ER background events observed during Run I $\gamma$ backgrounds originating from the material contamination in the detector components (internal source) and in the experimental hall (external source) were simulated, as well as the $^{39}$Ar $\beta$-decays. 
$^{39}$Ar $\beta$ electrons were generated according to the theoretical spectrum described by the phase space factors, the Fermi correction, as well as the first forbidden Gamow-Teller transitions (see \cite{Calvo:2016nwp} and its references). 

For the $\gamma$ background simulations radioactive decays ($\alpha$, $\beta$, electron capture etc.) of relevant sources, i.e. unstable isotopes in the $^{238}$U and $^{232}$Th decay series, $^{40}$K and $^{60}$Co, are simulated using the corresponding module of Geant4 \cite{Agostinelli:2002hh}. 
After a primary decay of the mother nucleus invoked in the simulation $\gamma$ photons are emitted in the radiative de-excitation process of the excited daughter nucleus, according to the decay branching ratios.
For the $^{238}$U and $^{232}$Th series a secular equilibrium is assumed. 
For the internal sources the primary particles (mother nuclei) are distributed homogeneously inside the material of each detector component. For the external sources an envelope surrounding the PE neutron shield is defined and the sources are distributed uniformly over its surface. 
The simulation data was generated separately for all detector components and radioactive sources, allowing later to scale each contribution according to the measured activities. 

\begin{figure}[hbt]
\centering
\includegraphics[width=0.49\textwidth]{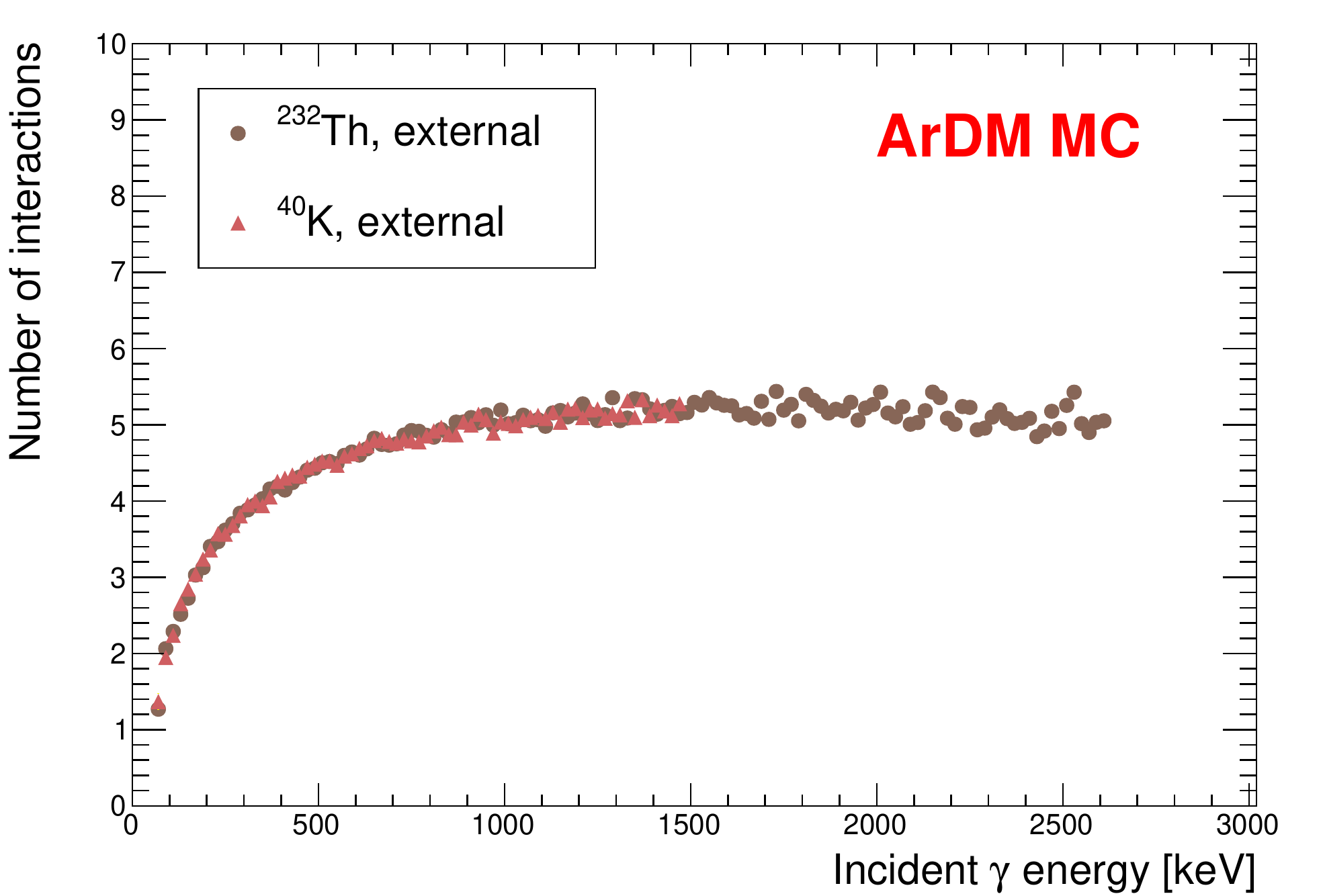}
\includegraphics[width=0.49\textwidth]{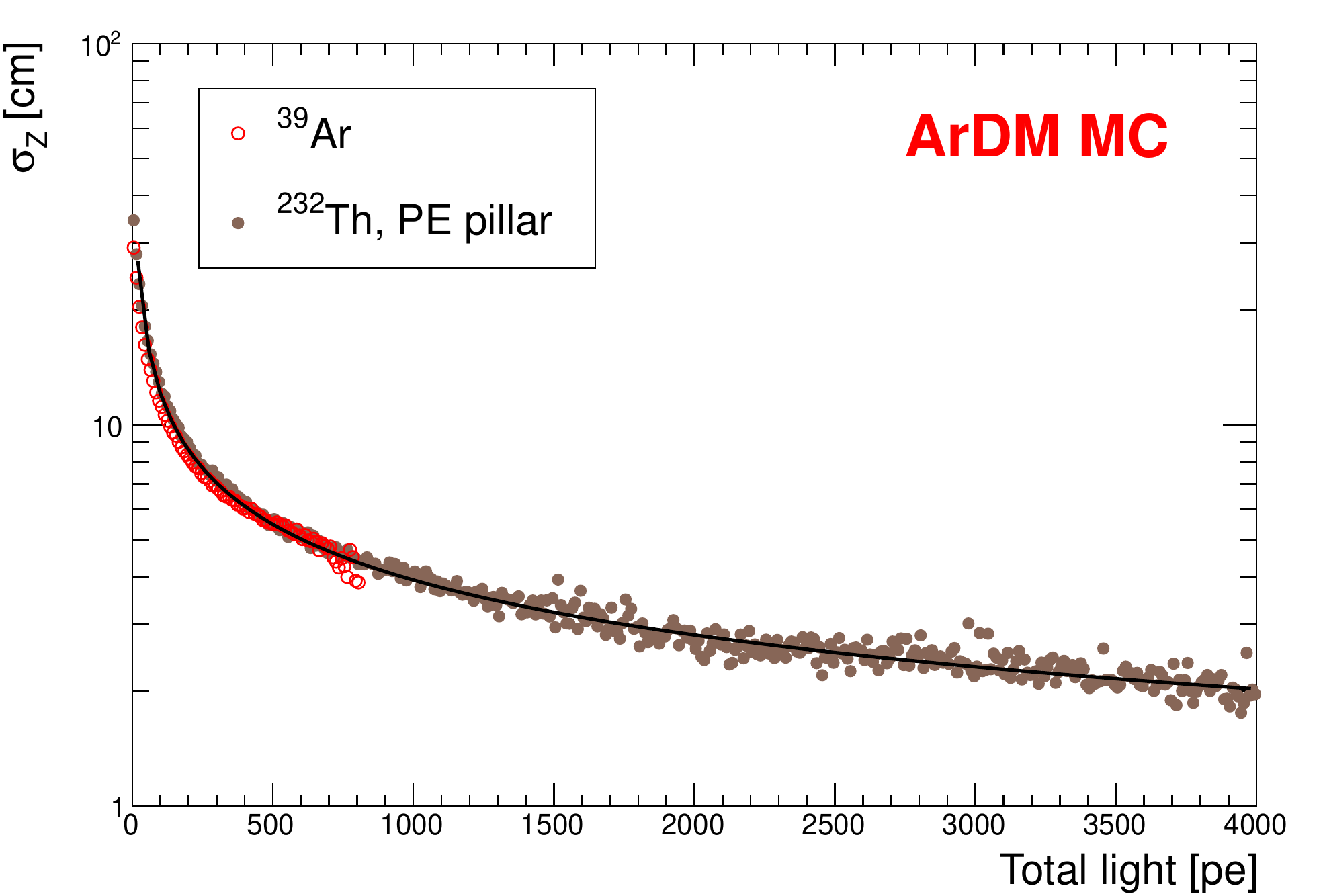}
\caption{Simulated characteristics of the ER background events in ArDM. Left: the mean number of interaction vertices (interaction multiplicity) in the LAr target for $\gamma$ events, as a function of the $\gamma$ energy at the incidence into the LAr volume. The values have been obtained from simulations of external sources, $^{232}$Th series (circle) and $^{40}$K (triangle). Right: The standard deviation of the true vertical positions of the interaction vertices, $\sigma_{Z}$,  or the centre of mass for multiple-scatter events, in the middle of the detector, $0.5 < TTR < 0.6$, as a function of the total detected light, for $^{39}$Ar-$\beta$ (open circle) and $^{232}$Th-series $\gamma$ (full circle) events from the polyethylene (PE) pillars of the drift cage. }
\label{fig:mc_gamma}
\end{figure}

Due to short mean free paths compared to the size of ArDM (6--20\,cm for the $\gamma$ energies 200--2600\,keV) most of the interaction vertices from $\gamma$ background events are found near the edge of the LAr target (self-shielding effect). 
More than 80\% of the $\gamma$ events contain two or more interaction vertices (Compton scatterings and photoelectric effect). 
As an example the mean interaction multiplicity of external $\gamma$ background events in the LAr target, as a function of the $\gamma$ energy at the incidence into the LAr volume, is shown in Fig. \ref{fig:mc_gamma} (left). 
The mean value of the multiplicity increases from $\sim$2 at a $\gamma$ energy of 100\,keV to $\sim$4.5 at 500\,keV, saturating at a value of $\sim$5 above 1\,MeV. In contrast, more than 99\% of the $^{39}$Ar events are single-scatter events, while the remaining events are multiple-scatter events from Bremsstrahlung.

The multiple-scatter $\gamma$ events are confined to a small volume. The standard deviation of the true vertical positions of the interaction vertices, $\sigma_{Z}$, in the middle of the detector, $0.5 < TTR < 0.6$, is shown as a function of the total detected light  
in Fig. \ref{fig:mc_gamma} (right), for $^{39}$Ar-$\beta$ and $^{232}$Th-series $\gamma$ events from the polyethylene pillars of the drift cage. The mean position deviation was found to be 3--4\,cm on average, and approximately constant above 500\,keV. 
This leads to a similar reconstructed vertical position resolution ($TTR$) for $^{39}$Ar $\beta$ and $\gamma$ events, despite the difference in scattering multiplicity of these processes. The energy dependence can approximately be explained by the statistical fluctuation of the binomial distribution, $\sigma_{Z} (L_{\rm tot}) = A \cdot \{L_{\rm tot} \cdot TTR \cdot (1 - TTR)\}^{\frac{1}{2}} + B$, with $TTR = 0.55$ and the fit parameters $(A, B) = (244, 0.13)$, the solid line plotted in the same figure. 
Averaging over the full energy range, weighted by the $\beta$ or $\gamma$ emission spectra, the resolution was found to be $\sim$9\,cm, which reduces to $\sim$6\,cm above 100\,$pe$, for both, $^{39}$Ar and $\gamma$ backgrounds.

\subsection{Electron recoil background sources}
\label{sec:gamma}
The observed ER events originate from \ar-\bet\ decays, radioactive contaminants in the materials surrounding the LAr volume  (internal sources) and in the experimental hall (external sources) including \ur\ , \tho\ , \pot\ and \cob\ series.

The radioactivity of the detector components was determined in a material screening campaign, using the high-purity germanium (HPGe) detector facility at LSC. Characteristic \gam\ lines from different nucleus were identified and the corresponding activities were estimated. Main contributions to the \gam\ activities were found to originate in the\ur\ and \tho\ series, and additionally in \pot\ or \cob. The detailed results were shown in Ref.\,\cite{1475-7516-2017-03-003}.

\begin{figure}[hbt]
\centering
\includegraphics[width=0.95\textwidth]{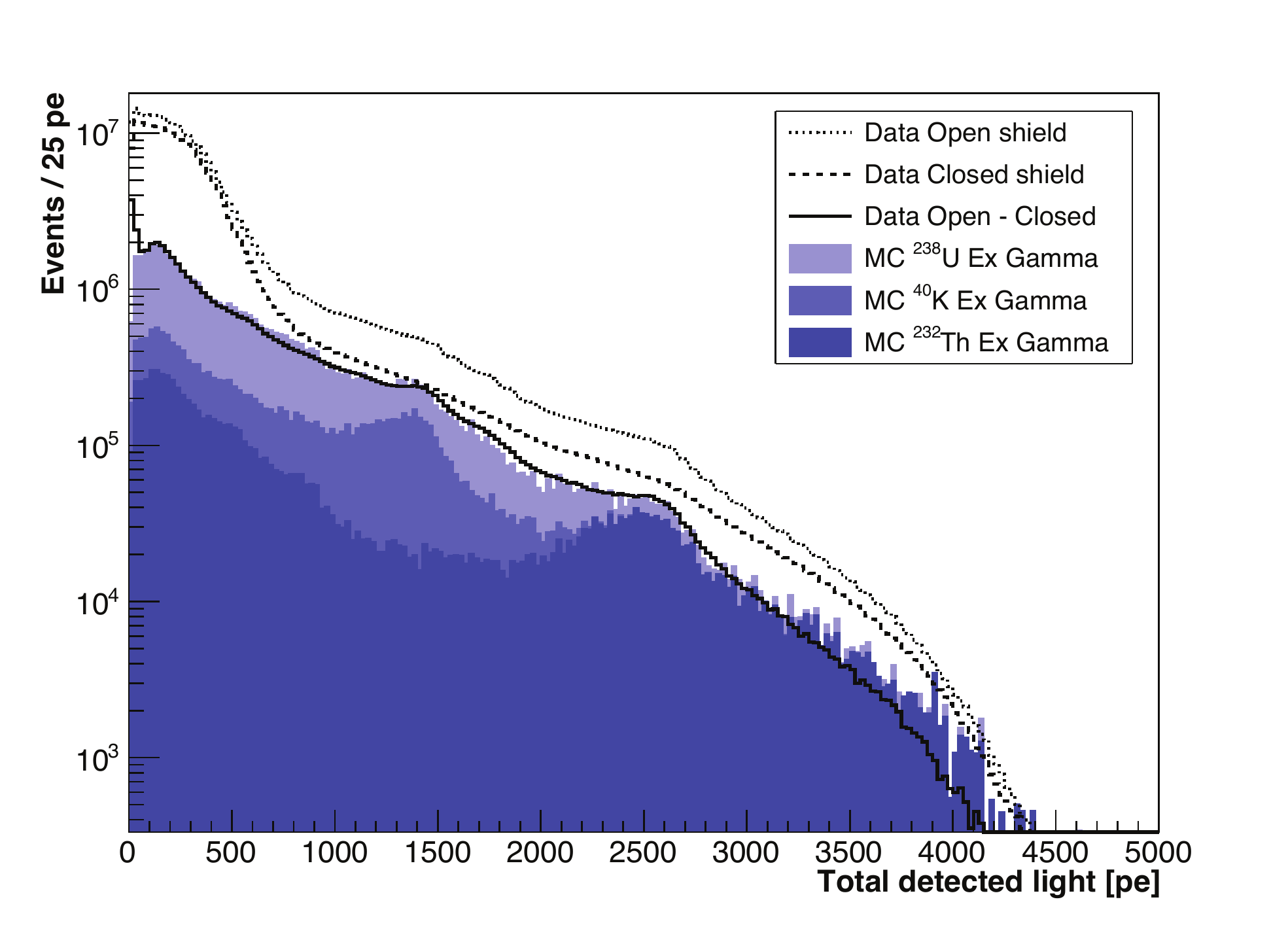}
\caption{The comparison between the external \gam\ background MC simulation (MC \ur\ , \tho\ and \pot\ ) and data using the difference of light spectra obtained from data (solid line) taken with open (dotted line) and closed (dashed line) top shield detector configurations. }
\label{fig:diff2}
\end{figure}

The detector vessel is surrounded by an equilateral octagon cylindrical polyethylene shield, where the top cap is a pre-assembled unit which can easily be opened and closed by crane. The external \gam\ flux are determined from \ardmrI\ data based on the difference in the measured ER events between the so-called open and closed top shield detector configuration. In order to model the difference in the light spectra between the open and closed shield configurations, it is assumed that the contribution to the difference is dominated by external \gam\ sources. The distribution of events as a function of $TTR$ vertical position value is shown in Figure\,\ref{fig:diff1} for the open and closed shield configurations, and with the difference of the two data. The difference of the total detected light spectrum of the open and closed shield configurations was found to be well described in the Monte Carlo model by a combination of the \gam\  spectra obtained from simulation of the \ur, \tho, and \pot\ radioactive series distributed homogeneously on an envelope around the \ardm\ detector, as shown in Figure\,\ref{fig:diff2}.

The number of events in the Monte Carlo spectrum for the source $i$ ($i = {\rm ^{238}U,\,^{232}Th,\,^{40}K})$ is given as $N_{{\rm MC}i} = f_i \cdot N_{\gamma i} \cdot \epsilon_i$, 
where $f_i$ is the scaling factor obtained from a fit to the light spectrum difference in the data, $N_{\gamma i}$ the number of generated $\gamma$ photons in the simulation and $\epsilon_i$ the event acceptance, which can be determined by the simulation. 
The equivalent external $\gamma$ flux $j_{\gamma i}$ originating from the source $i$ can then be obtained as 
\[
j_{\gamma i} = \frac{f_i \cdot N_{\gamma i}}{2 \cdot t_{\rm D} \cdot A} \ {\rm [cm^{-2}s^{-1}]}, 
\] 
where $t_{\rm D}$ [s] is the live time for the data taking and $A$ [cm$^2$] the total area of the surface of the envelope.
In this model spatially uniform and isotropic fluxes are assumed, as well as a secular equilibrium for the \ur\ and \tho\ decay series. The estimated external \gam\ flux result is $0.72$, $0.13$ and $0.05$ cm$^{-2}$\,s$^{-1}$ for \ur,  \tho\ decay chains, and \pot\, respectively.

The samples from the Monte Carlo simulations have been fitted to the data with a Bayesian model evaluation, using the Bayesian Analysis Toolkit (BAT) \cite{Caldwell20092197}. The MC template based binned likelihood model is a linear sum of the simulation samples obtained for the various radioactive isotopes in the light signal bins. The model assumes Gaussian uncertainties in each total detected light bin. Given the data, BAT finds the best parameters in the template likelihood model by numerically evaluating the Bayes theorem, varying the parameters of the template model, and calculating various credibility intervals from the full posterior distribution. The model parameters are the individual relative contributions of each Monte Carlo sample and the light yield scale. This latter parameter was introduced to allow for uncertainties in the light yield calibration in the data. Typically, less then $\sim$5\% variation was sufficient to describe any possible light scale uncertainty.

\begin{figure}[hbt]
\center
\includegraphics[width=0.9\textwidth]{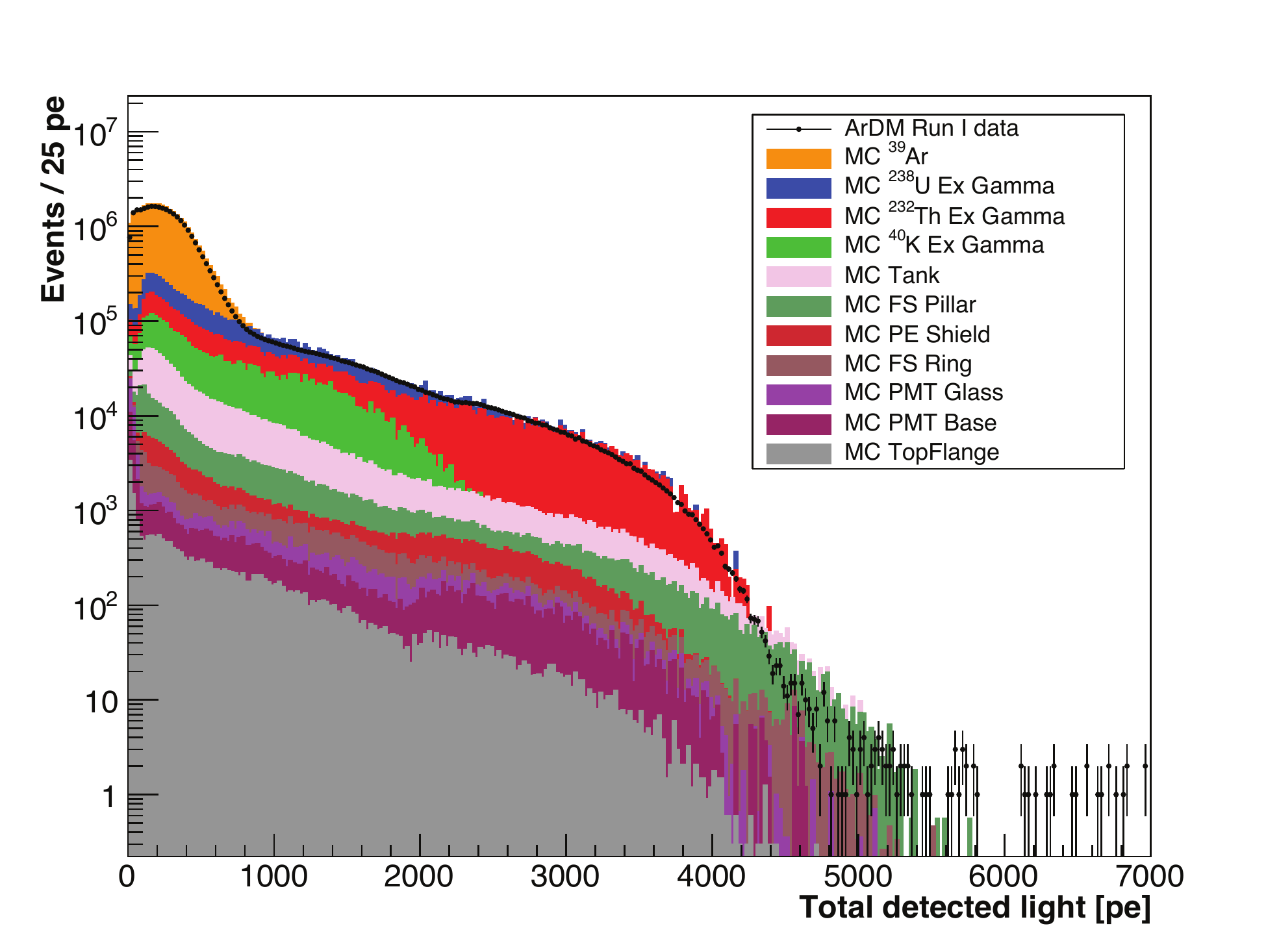}
\caption{Comparison of the experimentally obtained ER light spectrum to various backgrounds derived from the MC simulations. The events are selected using $0.4 < TTR < 0.6$ and $0 < f90 < 0.6$.}
\label{fig:gammaBKGSum}
\end{figure}

Figure\,\ref{fig:gammaBKGSum} shows the spectrum of the total detected light for ER events in data (black dots) compared to Monte Carlo simulations (filled coloured histograms). For this comparison a data sample of $\sim150$ million events was selected requiring $0.4 < TTR < 0.6$ and $0.0 < f90 < 0.6$. For the weighting of the internal detector components we use the activities obtained from the material screening results. For the external sources the results obtained from \gam\ flux evaluation are used, in particular the ratios of the fluxes of the external sources are constrained to be the same within a factor of two in both model evaluations. The \ar\; contribution is a free parameter during the fit. The level of agreement, shown in Figure\,\ref{fig:gammaBKGSum}, between MC simulation and data demonstrates the good understanding of the observed data and the different sources of events.

The dominant background component is \ar-\bet\ decays and shows a good agreement with the expected value of $\sim$1\,Bq/kg, amounting to $\sim$74\% of the selected events. The remaining contributions are compatible with events originating from internal and external sources, amounting to $\sim$22\% and $\sim$4\% of the  events, respectively. The composition of the various background sources is shown in Table \ref{table:breakdown} as rates. The uncertainties around the rates have been estimated by allowing a variation in the background contributions by 50\% and repeating the Bayesian evaluation. The \ur\ contribution was found to have the largest relative uncertainty, however, overall it only presents a $\sim$25\% error in the total external gamma rate. The estimated uncertainty on the total rate from Monte Carlo simulation is at a level of $\sim$7\%.

\begin{table}[htb]
\begin{center}
\vspace{2mm}
\begin{tabularx}{0.5\columnwidth}{p{60mm}@{\extracolsep{\fill}}cc} Component      &  Rate [Hz] \\
\hline
 \ar 	      &	219$\pm$4 	\\
 External \ur 	 & 27$\pm$16 	\\
 External \tho 	      &	 22$\pm$3     \\
 External \pot        & 17$\pm$8   \\
 Tank        & 6.6$\pm$2.9   \\
 Fieldshaper pillar  & 2.0$\pm$0.9    \\
 Polyethylene shield        & 0.6$\pm$0.3\\
 Fieldshaper ring              &    0.4$\pm$0.2 \\
 PMT glass          & 0.3$\pm$0.1 \\
 PMT base             & 0.3$\pm$0.1 \\
 Top Flange             & 0.2$\pm$0.1   \\
 \hline
 Total MC	& 	295 $\pm$19 \\ 
 Total data &   274 \\
\hline
\end{tabularx}
\caption{Electron recoil background composition in the Monte Carlo model. The events are selected using $0.4 < TTR < 0.6$ and $0 < f90 < 0.6$.}
\label{table:breakdown}
\end{center}
\end{table}

\subsection{Measurement of the \ar\ specific activity}
\label{sec:ar}
The  \ar\ specific activity can be determined from the model evaluation results presented in the previous section, using the formula $N_{\ar} = t_{\rm D} \cdot \epsilon \cdot m \cdot \mathcal{A} $. Here $N_{\ar}$ is the number of detected \ar\ events, $t_{\rm D}$ is the live time of the data taking, $\epsilon$ is the event acceptance determined from the Monte Carlo simulation, $m$ is the LAr mass and $\mathcal{A}$ is the \ar\ specific activity. The obtained \ar\ specific activity in \ardm\ detector is $\mathcal{A} = 0.95 \pm 0.05$ Bq/kg. The measurement uncertainties come from the fiducial volume estimation and the extracted fraction of \ar\ events from the analysis.

\section{Measurement of \rn\ emanation}
\label{sec:rn}
The internal \rn\ and its progeny dominate the detected events at medium energy in \ardmrI\; data. The emanation of \rn\ originates from the decay chain of the long-lived unstable isotope \ur\ being present in the detector components. As a noble gas with a 3.82~day half-life, \rn\ atoms can migrate from the source of their production into the active detector volume. Recoils produced by \alp\ particles from \rn\ and its progeny could mimic a signal expected from WIMP interactions and could be detected. To understand the radon background, we measured the emanation of \rn\ from detector components using data taken with GAr target at room temperature.

The signals from internal \rn\ and its progeny are monitored by counting the decays of the isotopes \rn, $^{218}$Po, and $^{214}$Po in the energy region between 5 and 7\mev, and the 5.49\mev\ \alp\ signals from the decay of \rn\ are used to measure the emanation rate of \rn. In order to identify the 5.49\mev\ \alp\ events originating from the bulk of the gas, a fixed energy window is used together with a $TTR$ discrimination, $0.32<TTR<0.64$, corresponding to 217.0$\pm5.7$\,liters of fiducial volume. To avoid the collisional quenching effect of argon triplet excimers due to the impurities present in GAr, we use detected light in the fast component between 500 and 800~$pe$ to select \rn\ events. 

\begin{figure}[H]
\begin{center}
\includegraphics[width=0.9\textwidth]{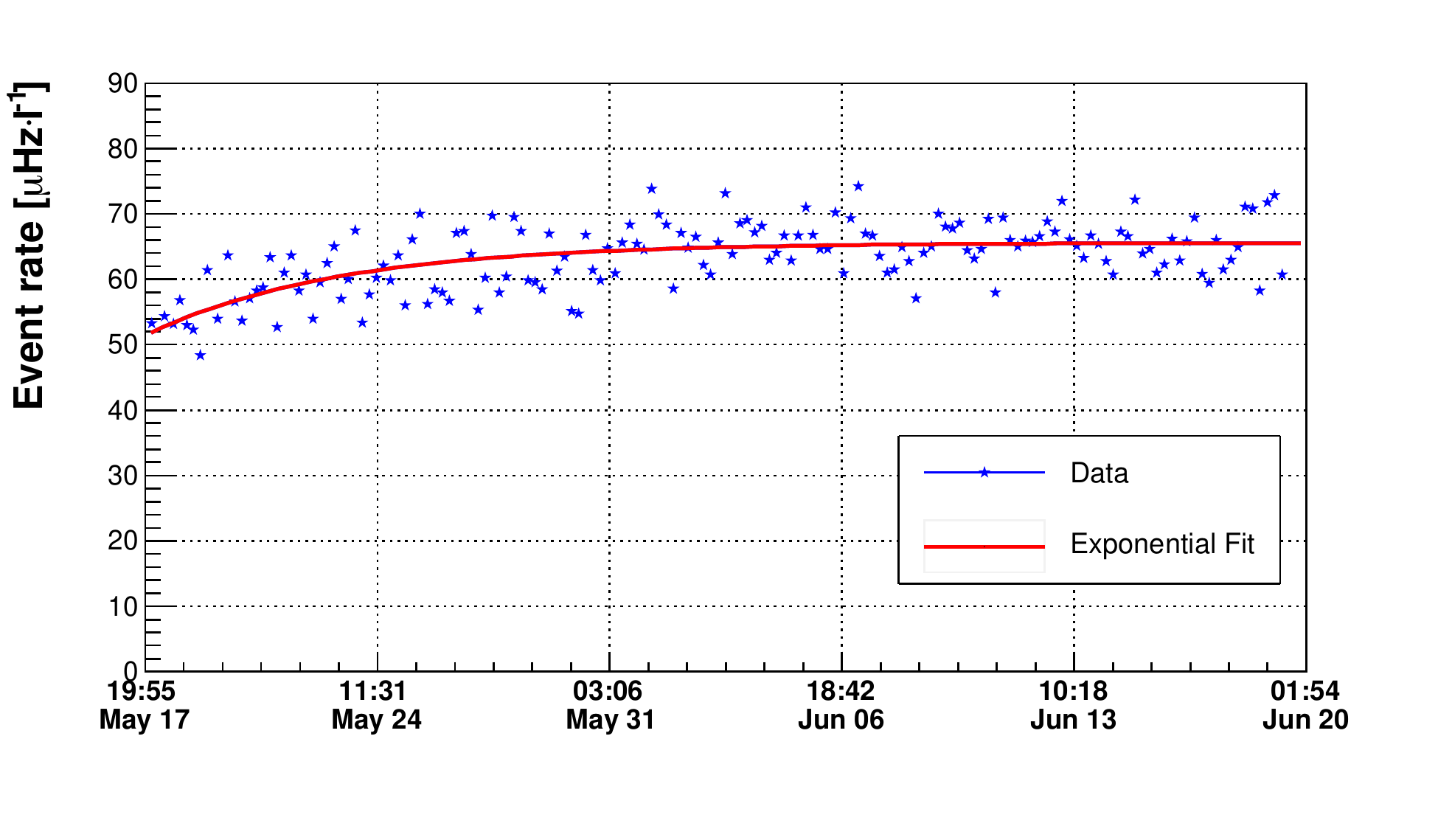}
\caption{The \rn\ rate as a function of time during data taking with room temperature GAr target. The red curve is an exponential fit to the data, $R(t) = A - B \times e^{-t/\tau}$, where the parameter  $\tau$ is fixed to $\tau_{1/2} = 3.82$~days.}
\label{fig:rnrate}
\end{center}
\end{figure}

Figure\,\ref{fig:rnrate} shows the \rn\ event rate variation after the detector tank was pumped and refilled with pure GAr. The red curve is an exponential fit to the data. During the fit the half-life was fixed to be 3.82~days according to the literature value. The \rn\ emanation rate increased exponentially, but not from zero. This indicates that the \rn\ atoms emanate from both the detector materials and the gas bottle. We set an upper limit for the emanation rate of \rn\ atom as 65.6$\pm$0.4\,$\mu$\hz/l for the \ardm\ detector.

\section{Large signal assessment of the ER spectrum}
\label{sec:Muon}

\begin{figure}[hbt]
\begin{center}
\includegraphics[width=0.9\textwidth]{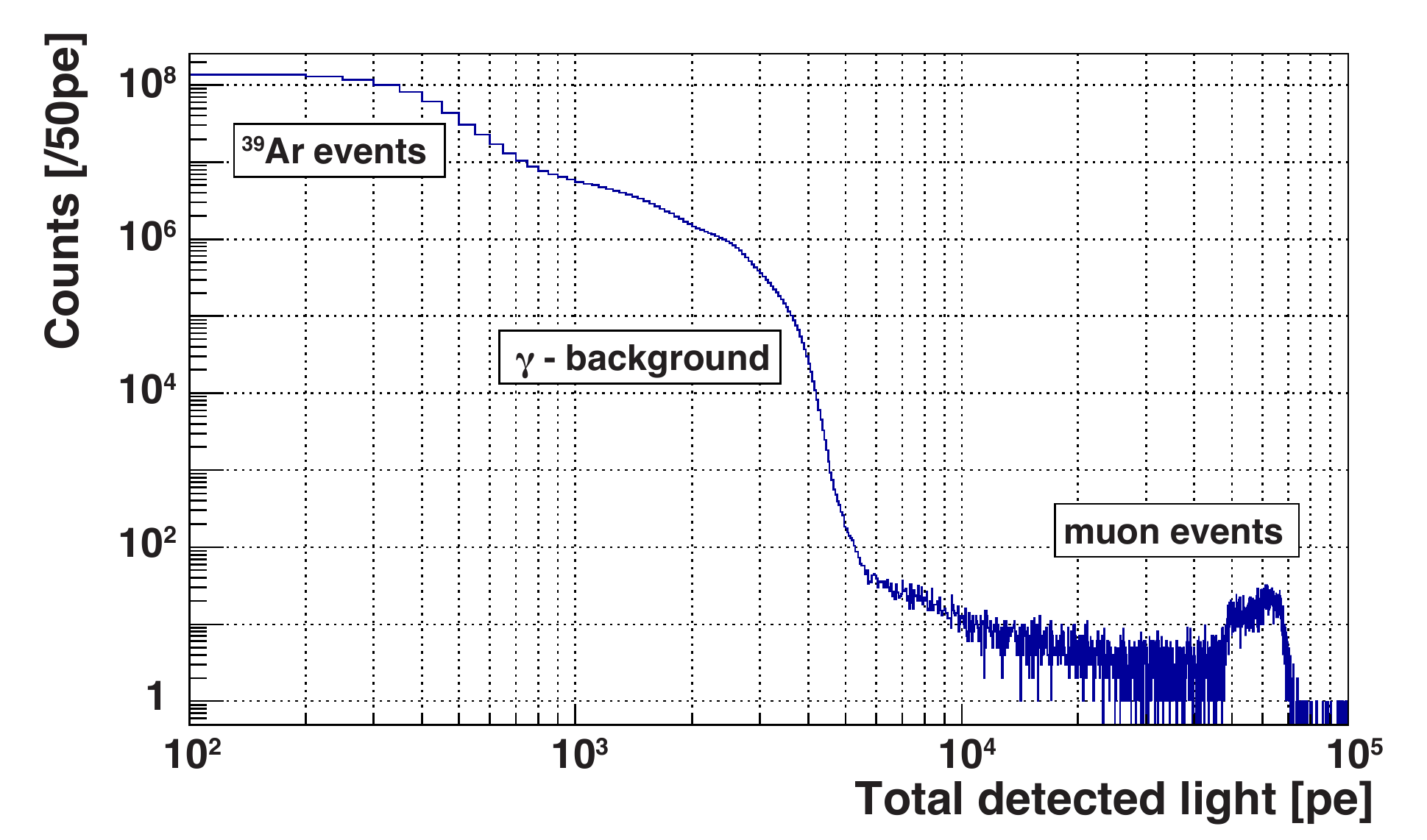}
\caption{Spectrum from 1.27$\times$10$^9$ ER events. The corresponding live time is 13.6\,d.}
\label{fig:Muonrate}
\end{center}
\end{figure}

For the analysis of the large signal ER spectrum a data set of 1.7$\times$10$^9$ events was used from the closed shield configuration. The events are selected requiring $f90$ to be around the centre of the ER band (0.26\,--\,0.55) leaving behind 1.27$\times$10$^9$ events which are shown in a double logarithmic histogram, in Figure\,\ref{fig:Muonrate}, spanning over 8 orders of magnitudes in the count rate. At low energies the spectrum is dominated by \ar\ $\beta$ events ($<10^3$\,$pe$), while the middle energy part is dominated by the general radiogenic \gam\ background, described in detail in the previous sections. We interpret events above the relatively sharp cut-off around 4$\times10^3$\,$pe$ as having been induced by cosmic muons. The tail and peak features in this part of the  spectrum are interpreted as showers in the rock, or muons crossing the active LAr target partially or fully. Some deformation of the spectral response has to be taken into account due to saturation effects in the analogue front-end or the ADC conversion, due to the large signal sizes. 

From the life-time of the detector during the recording of this data (1.18$\times$10$^6$\,s, 13.6\,days) and the amount of events under the main peak (7445) we find a cosmic muon rate in Hall\,A of LSC of 2.1$\times10^{-3}$\,m$^{-2}$\,s$^{-1}$, by assuming an effective area of 3\msq\ of the \ardm\ LAr target. If we include the events from the tail (in total 12318) we find a rate of 3.5$\times10^{-3}$\,m$^{-2}$\,s$^{-1}$. These values compare very well to the results from Ref.~\cite{Bettini:2012fu} of 2--4$\times10^{-3}$\,m$^{-2}$\,s$^{-1}$.

\section{Electron recoil statistical rejection power in Run I}
\label{sec:elike}
 One of the most important property of LAr as a target is the capability to reject electron recoil background processes at a very high level. This is possible due to the discrimination power of the pulse-shape from the LAr scintillation light. The classification of events into ER and NR-type allows to quantify the rejection capability of ArDM. The statistical rejection power for ER events is evaluated by calculating the leakage of events from ER ($0.2 < f90 < 0.6$) into the NR region ($0.7 < f90 < 1.0$). The ER contamination (${\cal ERC}$) is defined as the probability of incorrectly classifying an ER event as a NR event given a particular level of NR acceptance. The PSD rejection power (${\cal REJ}$) is then calculated from the tail integral of the ER Gaussian function around the NR region, as follows,
\begin{eqnarray}
\label{eq:REJ}
{\cal ERC}(N) &=& \frac{\int_{M_{\rm nr} - N \cdot \sigma _{\rm nr}}^{M_{\rm nr} + 3 \cdot \sigma _{\rm nr}} G_{\rm er}(M_{\rm er}, \sigma _{\rm er})}{\int_{0.0}^{M_{\rm nr} + 3 \cdot \sigma _{\rm nr}} G_{\rm er}(M_{\rm er}, \sigma _{\rm er})}\ ,   \nonumber \\
{\cal REJ}(N) &=& \frac{1}{{\cal{ERC}}(N)}  \ .
\end{eqnarray}

In the formula $N$ is the measure of the amount NR acceptance, expressed in Gaussian sigma units distance from the mean, $M_{\rm nr}$, of the Gaussian approximated NR event distribution, $M_{\rm er}$ and $\sigma _{\rm er}$ are the mean and sigma of the Gaussian functions, $G_{\rm er}(M_{\rm er}, \sigma _{\rm er})$, fitted to the data in the ER region.

\begin{figure}[hbt]
\begin{center}
\includegraphics[width=0.9\textwidth]{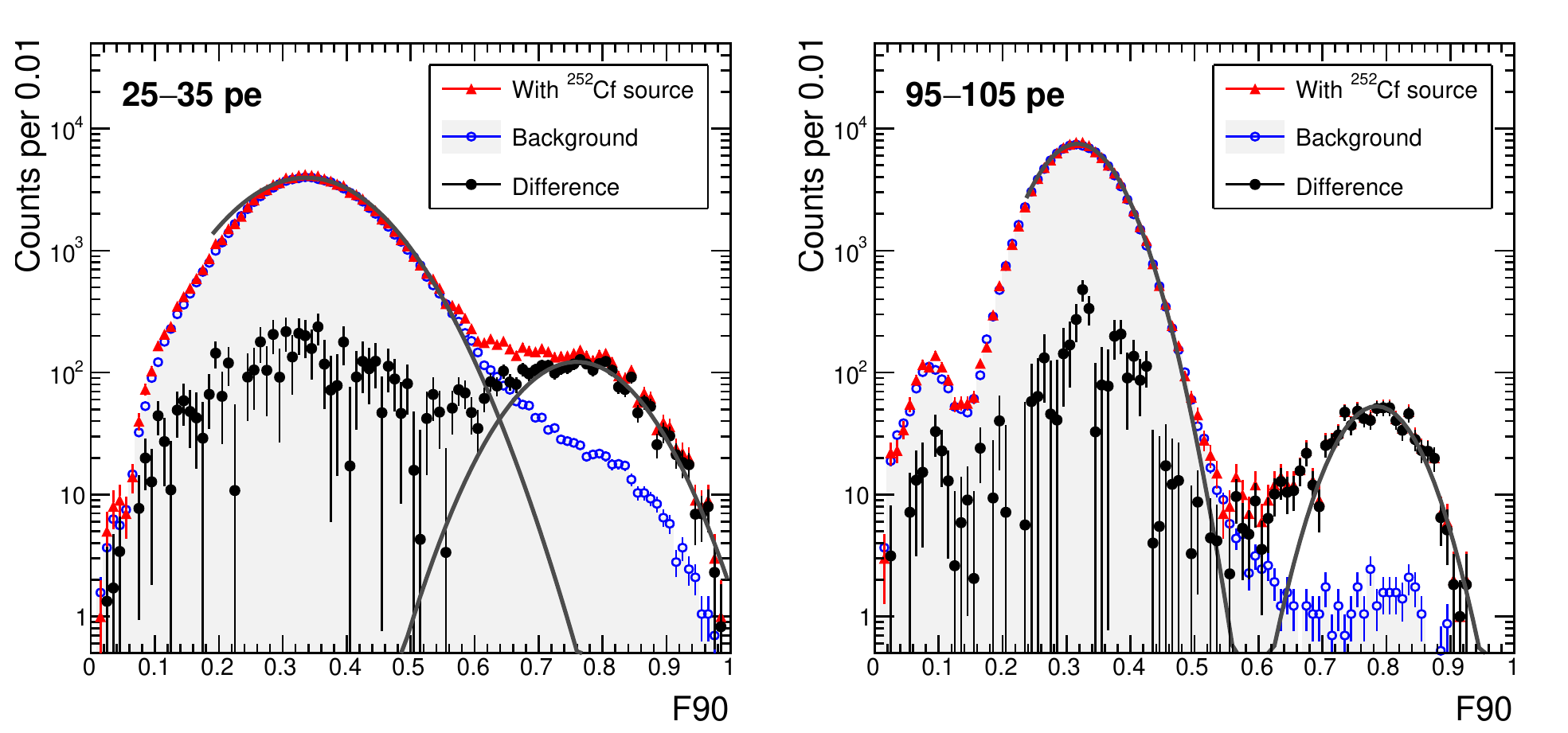}
\caption{Gaussian fit to $f90$-histograms for selected energy bins of $25-35\,pe$ (left) and $95-105\,pe$ (right). The figure shows the data with the \cf\ source (red triangles), the background data taken without the \cf\ source (blue dots) and the difference the two dataset (black dots). The ER and NR band fit results are shown as solid, gray lines.}
\label{fig:fitsample}
\end{center}
\end{figure}

\begin{figure}[hbt]
\begin{center}
\includegraphics[width=0.9\textwidth]{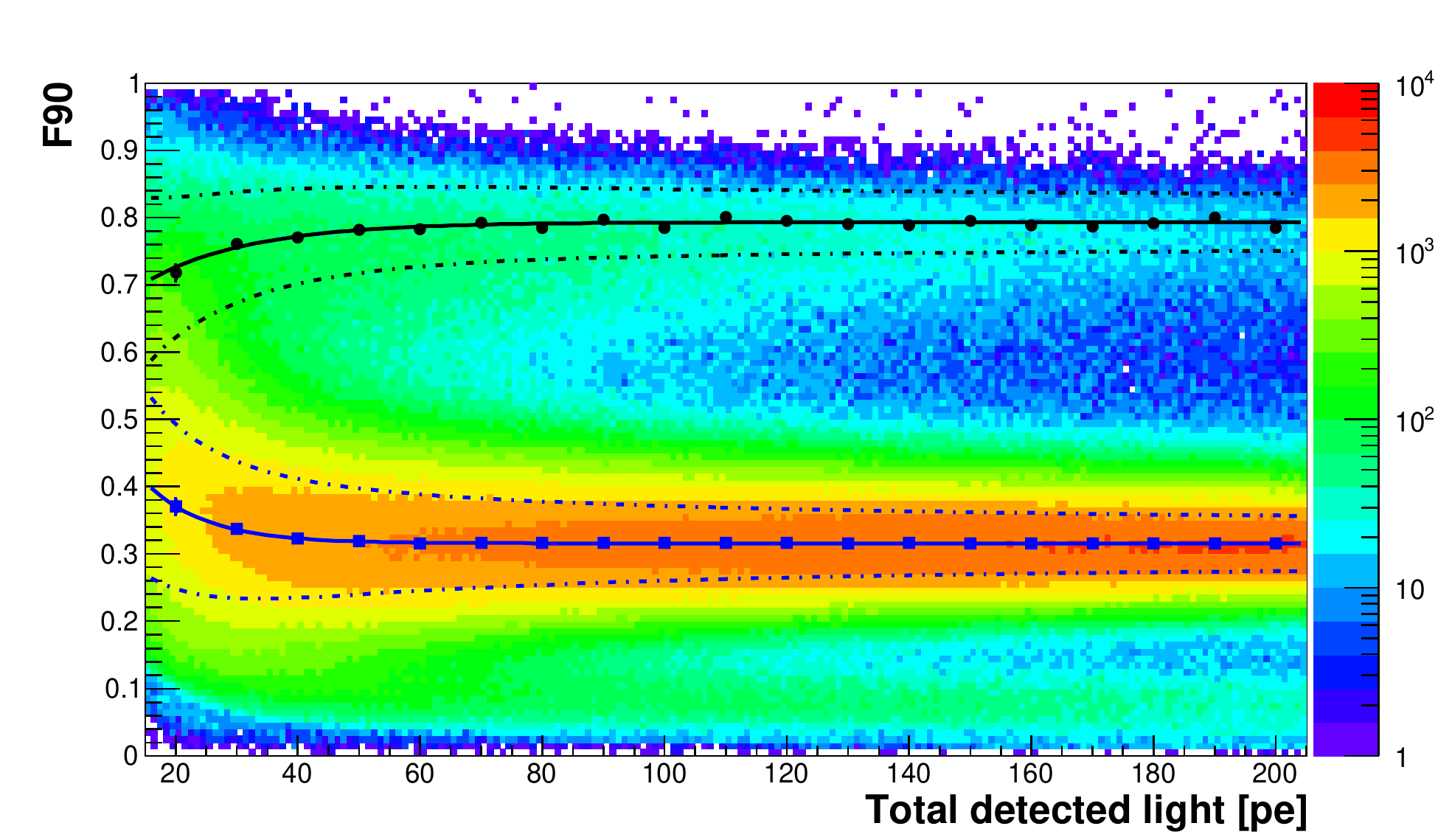}
\caption{$f90$ vs total detected light using $^{252}$Cf data. The black dots and blue squares show the $f90$ mean values for NR and ER events of each energy slice, respectively. The black and blue solid lines are the fitted parametrisations for the $f90$ mean values for NR and ER events, respectively. The black and blue dashed lines cover the 1\,$\sigma$ region around the obtained $f90$ mean values.}
\label{fig:speration}
\end{center}
\end{figure}

In this analysis a dataset of 23.8$\times$10$^6$ events is used. The data has been collected during \ardmrI\ with \cf\ calibration source placed around the detector. During the analysis $f90$ distributions are taken in various total light bins between 15 and 205\,$pe$. The distribution of events in the data is illustrated in Figure\,\ref{fig:fitsample}, which shows the number of events detected as a function of $f90$. The Gaussian functions in the ER region are fitted in the range $0.2 < f90 < 0.6$ to the data. In order to increase the signal-over-background ratio in the NR region we first subtract the background (dominated by ER events from \ar-\bet\ decays and the external \gam\ photons) from the $f90$ distributions and fit the Gaussian functions to each residual histogram in the range $0.7 < f90 < 1.0$. The result of the fitting procedure is also illustrated in Figure\,\ref{fig:fitsample}.
We parameterise the obtained $f90$ Gaussian mean values ($M_{\rm er}$, $M_{\rm nr}$), as a function the total light, with exponential functions. In a similar way, the $f90$ Gaussian $\sigma$-values ($\sigma _{\rm er}$,$\sigma _{\rm nr}$) are parameterised with a function of the form $\sigma (L_{\rm tot}) = A + B/(L_{\rm tot} - C)$. The fit results together with the parameterised mean and sigma values are shown in Figure\,\ref{fig:speration}.

\begin{figure}[hbt]
\begin{center}
\includegraphics[width=0.9\textwidth]{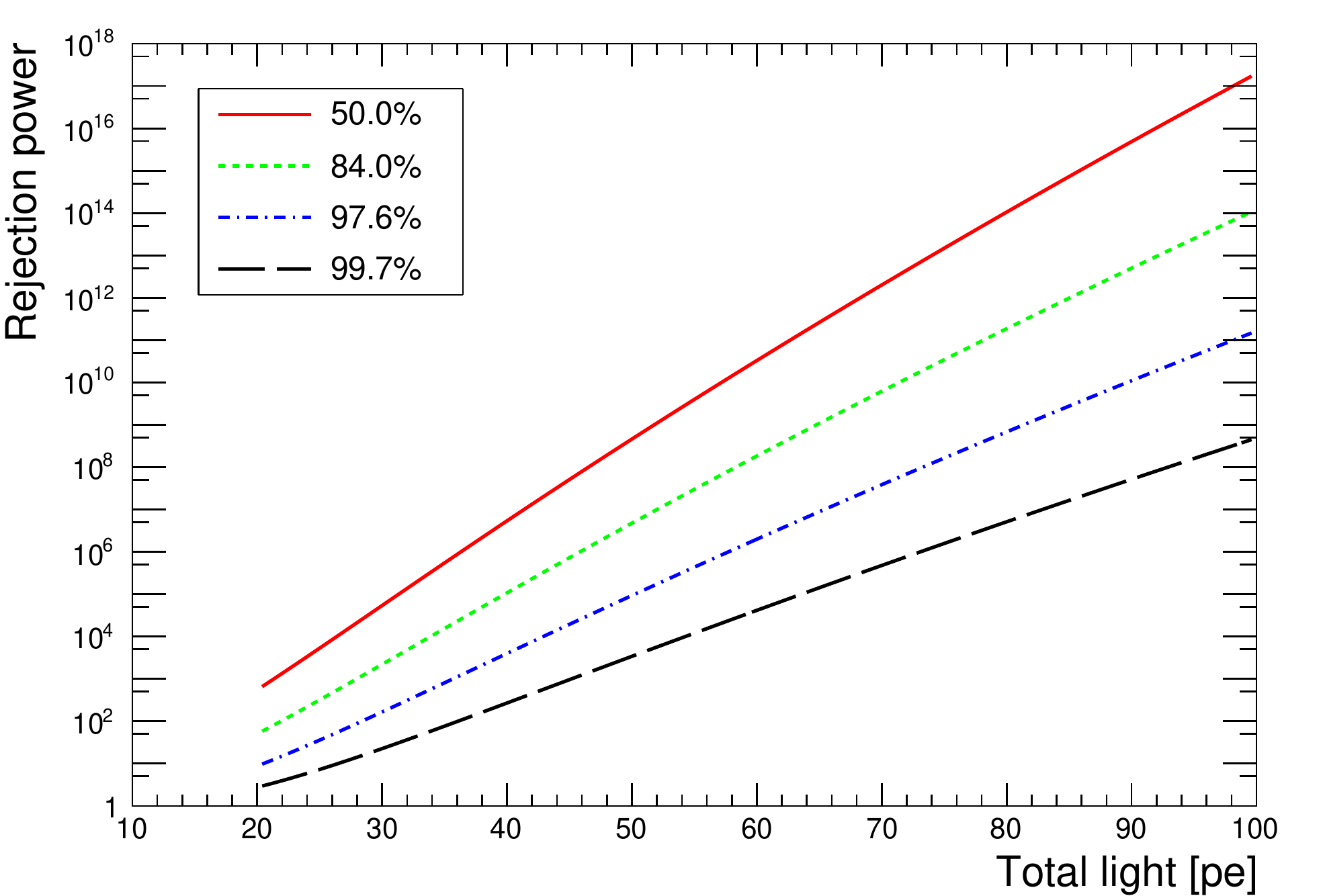}
\caption{The measured ER PSD rejection power for different NR events acceptance rates as a function of the detected number of photons.}
\label{fig:rej}
\end{center}
\end{figure}

The ER statistical rejection power is calculated from the parameterised values using Eq\,\ref{eq:REJ}. The results are shown in Figure\,\ref{fig:rej} as a function of the light signal for 50\%, 84.0\%, 97.6\%, and 99.7\% NR acceptance levels, corresponding to [0, 3$\sigma _{\rm nr}$], [-$\sigma _{\rm nr}$, 3$\sigma _{\rm nr}$], [-2$\sigma _{\rm nr}$, 3$\sigma _{\rm nr}$], and [-3$\sigma _{\rm nr}$, 3$\sigma _{\rm nr}$] limits across the calculated NR mean value $M_{\rm nr}$. As an example, for a 50\% NR acceptance level, the PSD rejection power is larger than 10$^8$ for events with total detected light larger than 50\,$pe$, corresponding to about 50\,keV$_{\rm ee}$ using the \ardmrI\ light yield. 

\section{Sensitivity to detect \ar\ decays in depleted argon}
\label{sec:depar}

Despite electronic recoil background can be suppressed by several orders of magnitudes by PSD, next generation LAr DM detectors with target sizes of several tens of tonnes, require argon depleted from radioactive impurities (DAr), above all from the $^{39}$Ar isotope, which is present in an abundance of about 1\,Bq per litre LAr, if produced from atmospheric argon (AAr). This is necessary to reduce the event rate in the target, as well as to suppress feedthrough into the signal region from background related to \ar\ beta decays. One of the cornerstones of the DS20k project is the provision of several tens of tons of depleted argon, collected from well gases. Further on a refinement of the argon in distillation columns is planned~\cite{Aalseth:2017fik}.  Presently about 150\,kg of DAr was produced and was recently qualified in the DS-50 detector. A depletion factor of 1/1400 was found with respect to atmospheric argon. 
It is not known and not proven that the depletion factor will be the same for all samples of gas extracted from underground wells, as local rock conditions can significantly alter the composition. Therefore, all future batches of DAr will have to be tested individually.
Thanks to the large target mass and the low background installation, the ArDM facility at LSC is ideally suited to determine the depletion factor for low \ar\ argon of future batches to be produced in view of DS-20k.

\begin{figure}[htbp]
\begin{center}
\includegraphics[width=0.36\textwidth]{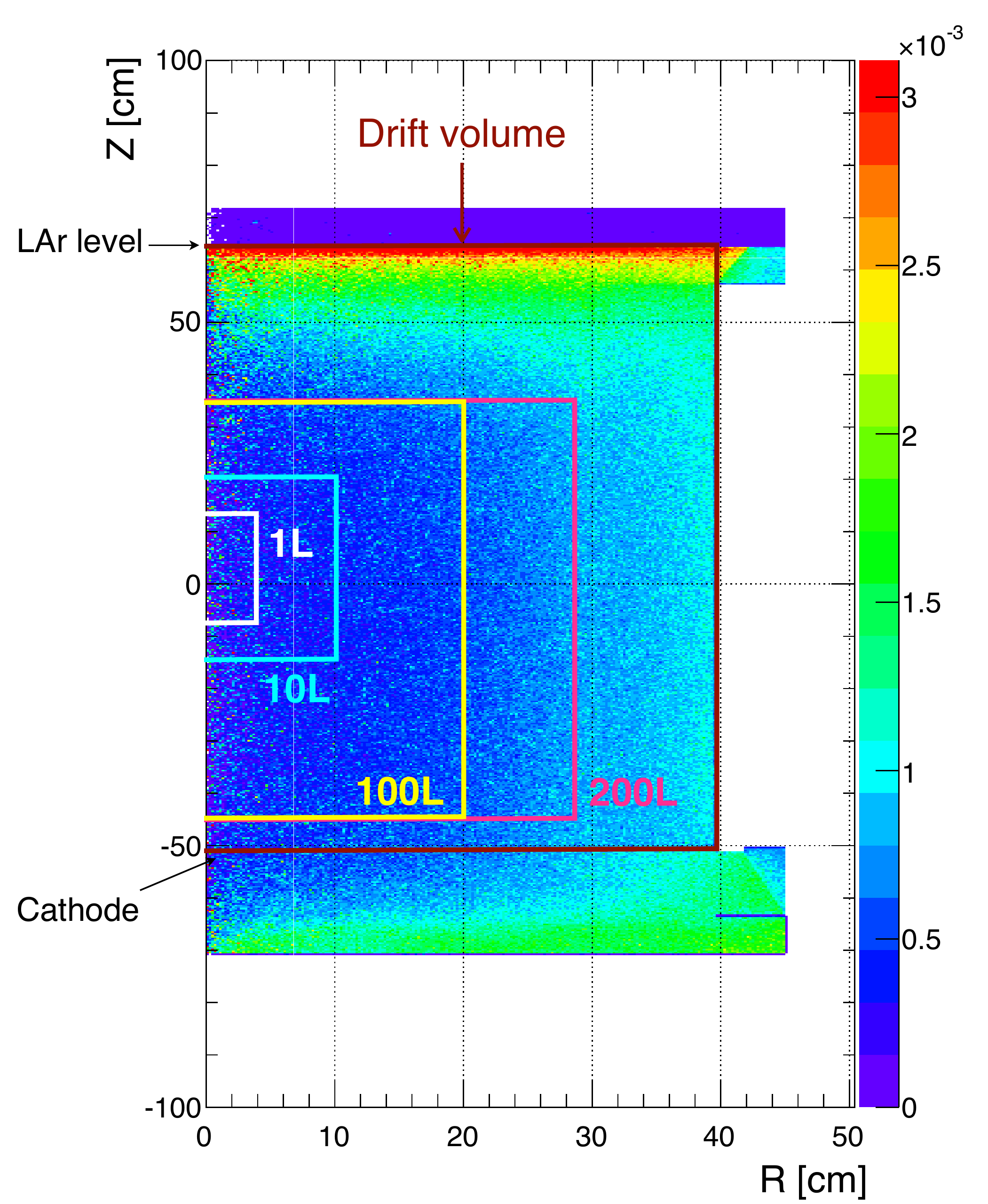}
\includegraphics[width=0.63\textwidth]{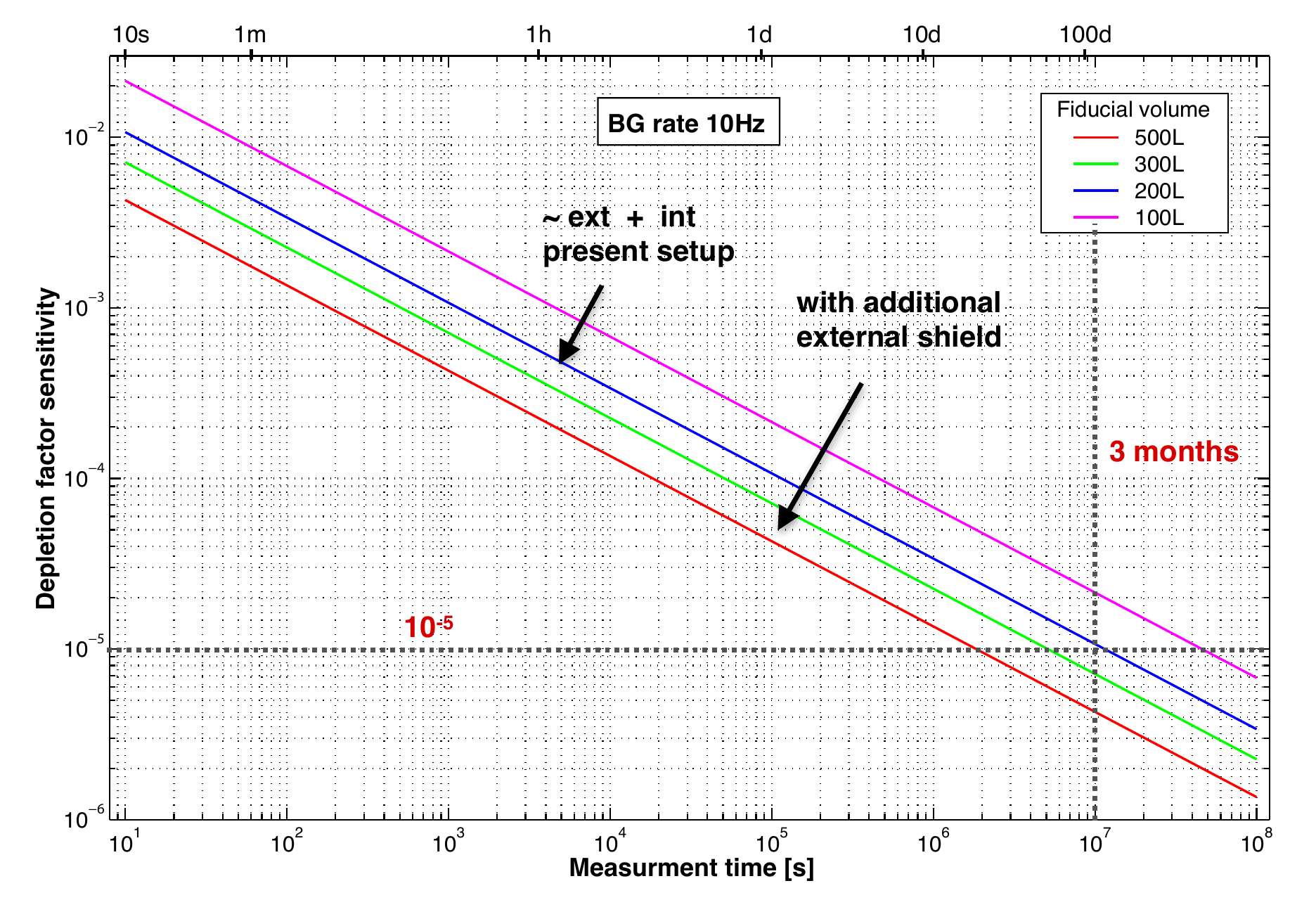}
\caption{Left: Simulated interaction vertices of the internal and external $\gamma$ backgrounds in the LAr target, plotted as a function of the vertical (Z) and the radial (R) coordinate. The boxes show the four different fiducial volumes under study, as well as the active drift volume of the ArDM TPC.
Right: 3$\sigma$ sensitivity to the \ar\ depletion factor at a given measurement time in ArDM.}
\label{fig:vertices_ar39sens}
\end{center}
\end{figure}

The ArDM sensitivity to the $^{39}$Ar $\beta$-decay activity in the LAr target was studied with respect to external and internal $\gamma$ backgrounds, based on the MC simulations as described in Section\,\ref{sec:ar}. For a first estimate the simulated interaction vertices of $\gamma$ photons in the LAr target are plotted in Fig.\,\ref{fig:vertices_ar39sens}\,(left) as a function of the vertical (Z) and the radial (R) coordinate. Contributions from all the simulated internal and external sources were scaled according to the measurements and then summed up. Thanks to the self-shielding of the large LAr target of ArDM, a clear accumulation of the events near the edges of the active volume is visible. Furthermore, more than 80\% of these $\gamma$ photons undergo multiple Compton scatters in the drift volume of ArDM. 
With an event vertex reconstruction based on S2 signals in the dual-phase mode, this allows an efficient reduction of the backgrounds inside a fiducial volume, using the outer volume as a veto. 

Four different fiducial volumes, depicted also in Fig.\,\ref{fig:vertices_ar39sens}\,(left), were investigated. An S2 threshold of 2\,keV for the energy deposit at each vertex was applied. The resulting rates of the $\gamma$ background for the four fiducialisations are summarised in Table\,\ref{table:fv_rate} with the rejection rate. The rates were calculated by the integral of the reconstructed light spectra in the range of the $^{39}$Ar $\beta$-spectrum, 50--600\,{\it pe}. 

\begin{table}[htb]
\begin{center}
\vspace{2mm}
\begin{tabularx}{\columnwidth}{c@{\extracolsep{\fill}}cccc} 
\hline
FV		& Total $\gamma$ Bkg in FV [Hz]	& Rejected [Hz]	& Remaining Bkg [Hz] 	& Rejection rate 	\\
\hline
200 $\ell$	& 32.7						& 21.6		& 11.1				& 66.0\% 			\\
100 $\ell$	& 11.3						& 9.5			& 1.8					& 84.5\%			\\
10 $\ell$		& 1.26						& 1.20		& 0.06				& 95.6\%			\\
1 $\ell$		& 0.164						& 0.160		& 0.004 				& 97.3\%			\\
\hline
\end{tabularx}
\caption{Simulated $\gamma$ background (Bkg) rates for four different fiducial volumes (FV) with and without event rejection based on the vertex reconstruction using S2 signals. The rates were calculated by the integral of the reconstructed light spectra (see Fig.\,\ref{fig:gammaBKGSum}) in the range 50--600\,{\it pe}.}
\label{table:fv_rate}
\end{center}
\end{table}

Figure\,\ref{fig:vertices_ar39sens}\,(right) shows the sensitivity of ArDM towards the \ar\ content in the target. The numbers are calculated for a 3$\sigma$ measurement of the event rate in the \ar\ $\beta$-spectrum over statistical background fluctuations.
These estimates  
indicate, that a measurement time of a few weeks in ArDM is needed to determine depletion factors up to 10$^5$ in respect to atmospheric argon, which is within the requirements for DS-20k. 

Until large amounts of depleted argon will be available, the DS-20k Collaboration envisages a first step to build and insert a smaller chamber (about 0.5$\ell$) into the center of the ArDM target, which can be filled with a separate line from the outside with already existing depleted argon from the DS collaboration. This chamber will contain a SiPM detector module in the top and in the bottom, and will also serve to get experienced with this new photo detection system. As DAr becomes available and more SiPM detectors are produced, a larger scale setup will be installed in ArDM at LSC to fully characterise the DAr batches needed for DS-20k.
 
\section{Conclusion}
\label{sec:conc}

In this paper measurement results have been presented on backgrounds and pulse shape discrimination of the \ardm\ experiment using Run I data. From the Monte Carlo model fitted to the data the ER background at low energy is found to be dominated by events originating from \ar-\bet\ decays, which contribute to the selected events at level of $\sim74\%$. In the $1 - 4$ MeV energy range the external \gam\ background dominates with an overall contribution of $\sim22\%$ and the internal detector components contribute at a level of $\sim4\%$ . The specific activity of \ar\ has been estimated to be $\sim1$Bq/kg. The external \gam\ background flux has been measured, using the light spectrum difference between open and closed top shield configurations, and the total flux is estimated to be $0.9$ cm$^{-2}$\,s$^{-1}$. From these measurements, we confirm the low background condition of \ardm\ detector.

At medium and high energies, signals produced by \rn\ and cosmic muons contribute to the data. From the data taken with GAr target at room temperature we estimate the \rn\ emanation rate of \ardm\ detector to be 65.6$\pm$0.4\,$\mu$\hz/l, which shows that the radon level was low during \ardmrI. From the ER light spectrum at high energy, the cosmic muon flux is estimated to be between 2.1 and 3.5 $\times10^{-3}$~m$^{-2}$~s$^{-1}$, which agrees well with independent measurements.

Using the pulse shape discrimination method the ER statistical rejection power is found to be more than 10$^{8}$ at $50$\% NR acceptance and more than 50~$pe$ detected. Together
with recent results reported by DEAP-3600~\cite{Amaudruz:2017ekt},
this study shows the excellent performance obtainable with liquid Argon targets at the tonne scale.

\section*{Acknowledgements}
We acknowledge the support of the Swiss National Science Foundation (SNF) and the ETH Zurich, as well as the Spanish Ministry of Economy and Competitiveness (MINECO) through the grants FPA2012-30811 and FPA2015-70657P.

We thank the directorate and the personnel of the Spanish underground laboratory {\it Laboratorio Subterr\'aneo de Canfranc} (LSC) for the support of the \ardm\ experiment. We also thank CERN for continued support of \ardm\ as the CERN Recognized RE18 project, where part of the R\&D and data analysis were conducted.

\bibliographystyle{JHEP}
\bibliography{ardmbib.bib}{}

\end{document}